\documentstyle[preprint,prl,aps]{revtex}

\title{ Excluded volume hadron gas model for
particle number ratios in $A$+$A$ collisions }

\author{ Granddon D. Yen,$^1$ Mark I. Gorenstein,$^{2,3}$ 
Walter Greiner$^4$ and Shin Nan Yang$^3$ }

\address{$^1$Institute of Physics, Academia Sinica, Taipei 11529, 
Taiwan \\
$^2$Bogolyubov Institute for Theoretical Physics, Kiev, Ukraine \\
$^3$Department of Physics, National Taiwan University, Taipei 10617, 
Taiwan \\
$^4$Institute for Theoretical Physics, Goethe University, 
Frankfurt, Germany }

\draft

\begin{document}

\maketitle

\begin{abstract}
We recapitulate a thermodynamically consistent excluded volume 
hadron gas model and examine its differences with other ``thermal 
models'' used in the literature.  Preliminary experimental data 
for particle number ratios in the collisions of Au+Au at the 
BNL\ \ AGS (11$A$ GeV/$c$) and Pb+Pb at the CERN\ \ SPS 
(160$A$ GeV/$c$) are analyzed.  For equal values of the hadron 
hard-core parameters the excluded volume model gives essentially 
the ideal gas predictions for the particle number ratios, which 
is similar to other thermal models.  We observe, however, the 
systematic excess of experimental pion abundances compared to 
the ideal gas results.  This effect can be explained in our 
model by a smaller pion hard-core volume compared to those of 
other hadrons.  The absolute values for particle number and 
energy densities at the chemical freezeout are predicted with 
a simultaneous fit to all these AGS and SPS particle number 
ratios.  
\end{abstract}


\section{Introduction}

Preliminary data for nucleus-nucleus (A+A) collisions with truly 
heavy beams have recently become available: Au+Au at 11$A$ GeV/$c$ 
at the BNL\ \ AGS and Pb+Pb at 160$A$ GeV/$c$ at the 
CERN\ \ SPS \cite{qm96}.  A systematic analysis of these data 
could yield clues to whether a short-lived phase with quark and 
gluon constituents, the quark-gluon plasma, exists during the 
hot and dense stage of these reactions.  

For the studies of matter properties in A+A collisions, it is 
of vital importance to determine whether local thermodynamical 
equilibrium in the system is reached.  Assuming such a local 
thermodynamical equilibrium at the final (freeze-out) stage of 
the process, one can calculate the particle number ratios 
without detailed knowledge of the very complicated dynamical 
evolution history of the system.  We remind that the 
{\em chemical} freezeout which determines the hadron number 
ratios does not necessarily coincide with the {\em thermal} 
freezeout which defines hadron momentum spectra.  

The aim of the present paper is to analyze the preliminary data 
for particle number ratios of Au+Au (AGS) and Pb+Pb (SPS) 
collisions within the framework of the thermodynamical 
equilibrium hadron gas model.  The name ``thermal model'' has 
often been used in the literature for this type of 
calculations.  We stress, however, that those ``thermal 
models'' used in the literature to calculate particle number 
ratios are, in fact, very different.  Therefore, our first step 
is to explain our model and clarify its difference from other 
versions of ``thermal model''.  

It seems natural to start with the ideal hadron gas at the 
freeze-out stage.  All known particles and resonances should be 
included in this gas and the resonance decay modes to the 
observed particles must be taken into account as well.  However, 
such an ideal gas model becomes inadequate in high-energy A+A 
collisions.  The chemical freeze-out parameters, temperature $T$ 
and baryonic chemical potential $\mu_b$, obtained from fitting 
the particle number ratios at AGS and SPS energies lead to 
artificially large values of total particle number densities 
which are much higher than any reasonable (or ``intuitively 
expected'') ones at the freezeout.  The total particle number 
density at the chemical freezeout within the ideal gas approach 
is $n_{tot}\approx 4\,n_o$ for the AGS and 
$n_{tot}\approx 8\,n_o$ for the SPS, where 
$n_o\cong 0.16$ fm$^{-3}$ is the normal nuclear density.  These 
numbers also exceed any experimental estimate obtained from the 
particle multiplicities and from the measurement of the volume 
of the system at the freezeout with pion-interferometry method.  
To suppress undesirable large values of particle number 
densities, the van der Waals (VDW) excluded volume procedure is 
used.  We will follow in the present paper the thermodynamically 
consistent excluded volume model of Refs.~\cite{gor91,gor93} 
which will be recapitulated in the next Section.  Some other 
``thermal models'' [4--8] also include VDW ``corrections'', but 
in some {\it ad hoc} and inconsistent ways.  It is often 
believed that the specific details of these VDW corrections are 
of minor importance.  This is, however, not the case.  The VDW 
repulsion does not essentially alter the particle number ratios, 
but it does always give a very strong suppression effect on the 
values of particle number densities themselves.  As we shall see 
below the total values of particle number densities are 
suppressed by a factor of 8 and 14 for AGS and SPS chemical 
freeze-out states, respectively.  Therefore, the influence of 
the VDW excluded volume procedure on particle densities, as well 
as on all other thermodynamical functions of the hadron gas, is 
very strong, and hence the correct form of the excluded volume 
model formulation should be employed.

\section{van der Waals Repulsion in the Grand Canonical Formulation}

For non-relativistic statistical mechanics, the use of a grand 
canonical ensemble is usually just a matter of convenience.  
However, in the hadron gas considered below it is unavoidable.  
In a relativistic theory one can not fix the number of particles, 
the number of pions in a hadron gas system, for example.  The 
average number of particles only makes sense in an equilibrium 
system, like in the case of a photon gas.  This average number 
is not conserved, but increases with increasing temperature of 
the system.  This takes place at all temperatures, even at very 
low ones when the thermal motion of each individual pion can be 
treated non-relativistically.  The canonical ensemble with a 
fixed number of pions has no physical meaning.  This statement 
is also valid, of course, for other hadrons.  Particle chemical 
potentials, in general, regulate not particle numbers but the 
values of conserved charges.  Strongly interacting matter has 
three conserved charges, viz., baryonic number $B$, strangeness 
$S$ and electric charge $Q$ (strangeness is conserved as we 
neglect ``slow'' weak interactions).  Therefore, the canonical 
ensemble can only be defined with fixed values of $B$, $S$ and 
$Q$, not the fixed numbers of pions, nucleons and other 
hadrons.  

Let's start with the ideal gas of one particle species with 
temperature $T$, chemical potential $\mu$ and volume $V$.  
The pressure $p^{id}$ is related to the grand canonical 
partition function ${\cal Z}^{id}$ and in thermodynamical limit 
$V\rightarrow \infty$, it has the following expression: 
\begin{equation}\label{idpress}
p^{id}(T,\mu) \equiv T\lim_{V\rightarrow \infty}
\frac{\ln{\cal Z}^{id}(T,\mu,V)}{V} = \frac{d}{6\pi^{2}}~
\int_{0}^{\infty}dk~ \frac{k^4}{(k^2+m^2)^{1/2}}~ f(k), 
\end{equation}
with 
\begin{equation}\label{distr}
f(k)~=~\left[\exp\left(\frac{(k^{2}+m^{2})^{1/2}~-~
\mu}{T}\right)~+~\eta ~\right]^{-1}
\end{equation}
being the momentum distribution function.  $d$ is the number of 
particle internal degrees of freedom (degeneracy) and $m$ is the 
mass.  The value of $\eta$ is $-1$ for bosons, $+1$ for fermions 
and $\eta=0$ gives the classical (Boltzmann) approximation.  
The ideal gas particle number density is given by 
\begin{equation}\label{idpart}
n^{id}(T,\mu)~
\equiv ~ \left(\frac {\partial\,p^{id}(T,\mu)}
{\partial \mu}\right) _{T} ~=~
\frac{d}{2\pi^{2}}~\int_{0}^
{\infty}dk~k^{2}~f(k)~.
\end{equation}
Furthermore, the entropy density is defined as 
$s^{id}=(\partial p^{id}/\partial T)_{\mu}$.  The energy density 
can be found from another thermodynamical relation 
$\varepsilon^{id}=Ts^{id}-p^{id}+\mu\,n^{id}$ and is written as 
\begin{equation}\label{ideps}
\varepsilon^{id}(T,\mu)~=~\frac{d}{2\pi^{2}}~
\int_{0}^{\infty}dk~k^2~(k^2+m^2)^{1/2}~f(k)~. 
\end{equation}

For a fixed particle number $N$, the VDW excluded volume 
procedure means the substitution of the volume of the system 
$V$ by $V-vN$, where $v$ is the parameter corresponding to the 
proper volume of the particle.  Note that this VDW procedure, 
interpreted in statistical mechanics as the gas of hard-sphere 
particles with radius $r$, requires that the volume parameter 
$v$ equal to the ``hard-core particle volume'', 
$\frac{4}{3}\pi r^3$, multiplied by a factor of 4 \cite{lan75}.  
To introduce the excluded volume ({\it \'{a} la} van der Waals) 
in our grand canonical ensemble formulation, we start with the 
volume substitution of $V$ by $V-vN$ in the canonical partition 
function $Z$ for each fixed $N$ separately.  The grand canonical 
partition function for the ideal gas system 
\begin{equation}\label{gcidpart}
{\cal Z}^{id}(T,\mu,V)~=~\sum_{N =0}^{\infty}
\exp\left(\frac{\mu N}{T}\right)~Z^{id}(T,N,V)
\end{equation}
then becomes 
\begin{equation}\label{gcpart}
{\cal Z}(T,\mu,V)~=~\sum_{N =0}^{\infty}
\exp\left(\frac{\mu N}{T}\right)~Z^{id}(T,N,V-v N)~
\theta(V-v N)~.
\end{equation}
There is a difficulty in the evaluation of the sum over $N$ 
in Eq.~(\ref{gcpart}) because of the $N$-dependence of the 
``available volume'', $V-vN$.  To overcome it, we perform a 
Laplace transform on Eq.~(\ref{gcpart}) \cite{gor91} and obtain 
\begin{eqnarray}\label{partlap}
\widehat{{\cal Z}}(T,\mu, x)~ & \equiv & ~
\int_{0}^{\infty} dV~\exp(-xV)~{\cal Z}(T,\mu,V) \\
\nonumber
& = &  \int_{0}^{\infty} d\widehat{V}~
\exp(-x\widehat{V})~{\cal Z}^{id}(T,\widehat{\mu},\widehat{V})~,
\end{eqnarray}
where $\widehat{\mu} \equiv \mu-vTx$ and 
$\widehat{V} \equiv V-vN$.  The second equality in 
Eq.~(\ref{partlap}) can be easily understood as follows.  We 
substitute ${\cal Z}(T,\mu,V)$ by the infinite sum of 
Eq.~(\ref{gcpart}), change the integration variable $V$ to 
$\widehat{V}$ in each term of the series and then use 
Eq.~(\ref{gcidpart}) to sum up the series again.  

From the definition of the pressure function 
\begin{equation}\label{pressdef}
p(T,\mu) \equiv T\lim_{V\rightarrow \infty}
\frac{\ln{\cal Z}(T,\mu,V)}{V}
\end{equation}
one concludes that the grand canonical partition function 
of the system, in the thermodynamical limit, approaches 
\begin{equation}\label{vlim}
{\cal Z}(T,\mu,V)| _{V\rightarrow \infty}~\sim~
\exp\left[\frac{p(T,\mu)~V}{T}\right]~. 
\end{equation}
From the first equality in Eq.~(\ref{partlap}) one sees that 
this exponentially increasing part of ${\cal Z}(T,\mu,V)$ 
generates an extreme right singularity in the function 
$\widehat{{\cal Z}}(T,\mu,x)$ at some point $x^*$.  For $x<p/T$ 
the integration over $V$ for $\widehat{{\cal Z}}(T,\mu,x)$ 
diverges at its upper limit.  Therefore, the extreme right 
singularity of $\widehat{{\cal Z}}(T,\mu,x)$ at $x^{*}(T,\mu)$ 
gives us the system pressure, 
\begin{equation}\label{presVDW}
p(T,\mu)~=~T~x^{*}(T,\mu)~. 
\end{equation}
Note that this direct connection of the extreme right 
$x$-singularity of $\widehat{{\cal Z}}$ to the asymptotic 
behavior $V \rightarrow \infty$ of ${\cal Z}$ is a general 
mathematical property of the Laplace transform.  Using the 
above equations we find 
\begin{equation}\label{ssingul}
x^{*}(T,\mu)~=~\lim_{\widehat{V} \rightarrow \infty} \frac{\ln
{\cal Z}^{id}(T,\tilde{\mu},\widehat{V})}{\widehat{V}}~,~~
\tilde{\mu}_{i} \equiv \mu ~-~
v~Tx^{*}(T,\mu)~.
\end{equation}
With Eqs.~(\ref{idpress}) and (\ref{presVDW}) we find from 
Eq.~(\ref{ssingul}) the transcendental equation for the pressure 
$p(T,\mu)$ of the gas with VDW repulsion in the grand canonical 
ensemble: 
\begin{equation}\label{vdweos}
p(T,\mu)~=~p^{id}(T,\tilde{\mu})~,~~~
\tilde{\mu}~ \equiv~ \mu ~ - ~v~ p(T,\mu)~. 
\end{equation}

At this point some comments are appropriate.  First, we remind 
that the singularity  $x^{*}$ of the 
$\widehat{{\cal Z}}(T,\mu,x)$ function has nothing to do with 
phase transition singularities of the system.  A singularity at 
$x^*(T,\mu)$ exists for the ideal gas as well --- one can easily 
recover the ideal gas formulae by putting $v=0$ in the above 
equations.  Another result which can puzzle a reader is the 
difference between the familiar form of the VDW repulsion, 
\begin{equation}\label{VDW}
p~(V~-~vN)~=~N~T~, 
\end{equation}
and our Eq.~(\ref{vdweos}).  According to Eq.~(\ref{VDW}) the 
ideal gas pressure, $p^{id}=NT/V$, increases due to the excluded 
volume repulsion as expected intuitively while Eqs.~(1) and 
(\ref{vdweos}) clearly shows that the ideal gas pressure 
decreases with the inclusion of the excluded volume repulsion.  
To resolve this ``paradox'', one should keep in mind the 
important difference between standard VDW gas treatment with 
Eq.~(\ref{VDW}) for fixed particle number $N$ and our system with 
fixed chemical potential $\mu$ in the grand canonical ensemble.  
To clarify this point, let us calculate the particle number 
density: 
\begin{equation}\label{partden}
n(T,\mu)~
\equiv ~ \left(\frac {\partial p(T,\mu)}
{\partial \mu}\right) _{T} ~=~
\frac{n^{id}(T,\tilde{\mu})}
{1~+~ v~n^{id}(T,\tilde{\mu})}~. 
\end{equation}
To have a direct comparison with Eq.~(\ref{VDW}) we now 
make the Boltzmann approximation, i.e., with $\eta =0$ in 
Eq.~(\ref{distr}).  We shall see later that at the AGS and SPS 
chemical freezeout, the Boltzmann approximation leads to a 
very good agreement (only a few percent deviations) for all 
thermodynamical functions with those calculated with exact Bose 
and Fermi distributions.  In Boltzmann approximation, the 
distribution function $f(k)$, as well as all thermodynamical 
functions of the ideal gas, possesses a momentum-independent 
factor $\exp(\mu/T)$.  One can easily find that 
\begin{equation}\label{boltz}
p(T,\mu)~=~\exp\left(-~\frac{v~p(T,\mu)}{T}
\right)~p^{id}(T,\mu)~,~~~~p^{id}(T,\mu)~= ~T~ n^{id}(T,\mu)~. 
\end{equation}
After some simple algebra, Eqs.~(\ref{partden},\ref{boltz}) give 
\begin{equation}\label{vdvmu}
p(T,\mu)~[1 - v\,n(T,\mu)] = n(T,\mu)\, T.  
\end{equation}
The form of Eq.~(\ref{vdvmu}) amazingly coincides with that of 
Eq.~(\ref{VDW}), and therefore, we have no contradiction with the 
standard VDW gas physics.  The crucial difference is that we have 
in Eq.~(\ref{vdvmu}) a $n(T,\mu)$ function instead of a fixed 
number $N$ (or a fixed particle density $N/V$) as is the case 
in Eq.~(\ref{VDW}).  In the grand canonical ensemble with fixed 
temperature and chemical potential, the VDW repulsion leads to a 
strong suppression of the particle number density.  It is just 
this suppression on the number of particles which leads to the 
decrease of the ideal gas pressure after the VDW repulsion begins 
to come into effect.  Eqs.~(\ref{vdvmu}) and (\ref{VDW}) have 
similar form, but their physical meaning are very different.  
Eq.~(\ref{VDW}) calculates the pressure as the function of $N$, 
$V$, $T$ while Eq.~(\ref{vdvmu}) gives particle number density 
$n(T,\mu)$, but only after Eq.~(\ref{boltz}) for $p(T,\mu)$ 
function is solved.  

With $p(T,\mu)$ as the solution of Eq.~(\ref{vdweos}) the 
particle number density is given by Eq.~(\ref{partden}).  
The entropy and energy densities are 
\begin{equation}\label{entrden}
s(T,\mu)~\equiv~\left(\frac{\partial p(T,\mu)}{\partial T}
\right)_{\mu}~=~
\frac{s^{id}(T,\tilde{\mu})}
{1~+~ v~n^{id}(T,\tilde{\mu})}~, 
\end{equation}
\begin{equation}\label{epsden}
\varepsilon(T,\mu)~\equiv ~ Ts ~-~ p~ +~\mu\,n~=~
\frac{\varepsilon^{id}(T,\tilde{\mu})}
{1~+~ v~n^{id}(T,\tilde{\mu})}~. 
\end{equation}
Eqs.~(\ref{partden},\ref{entrden},\ref{epsden}) reveal two 
suppression effects on the particle number, entropy and energy 
densities because of the VDW repulsion, namely 
\begin{itemize}
\item[{1).}] The modification of the chemical potential 
$\mu \rightarrow \tilde{\mu}$ as given in Eq.~(\ref{vdweos}).  In 
Boltzmann approximation it just leads to an additional factor 
$\exp(~-~v~p/T)$.  
\item[{2).}] A suppression factor 
$[1~+~ v~n^{id}(T,\tilde{\mu})]^{-1} < 1$.  
\end{itemize}

For a ideal gas system of several particle species $i=1,...,h$, 
the thermodynamical functions are additive and equal to the sums 
of their partial values for different particle species: 
\begin{equation}\label{idtot}
p^{id}(T,\mu_1, ..., \mu_h) =
\sum_{i=1}^h p_{i}^{id}(T,\mu_i)~,~~
\end{equation}
and similar expressions for $s^{id}$, $n^{id}$ and 
$\varepsilon^{id}$.  Note that the index $i$ includes all the 
information about the $i$-th particle, $m_i$, $d_i$, $\eta_i$, 
$\mu_i$, etc.  

The extension of the excluded volume procedure for several 
particle species is straightforward.  The grand canonical 
partition function of the ideal gas equals to the product of 
${\cal Z}^{id}_i(T,\mu_i,V)$ for each particle species ``$i$''.  
The excluded volume grand canonical partition function for 
several particle species, $i=1,...,h$, with proper volumes 
$v_1,...,v_h$ can then be written as
\begin{equation}\label{gcparth}
{\cal Z}(T,\mu_1,...,\mu_h,V)=
\sum_{N_1=0}^{\infty}...
\sum_{N_h=0}^{\infty}~\prod_{i=1}^{h}
\exp\left(\frac{\mu_i N_i}{T}\right)~Z_i^{id}(T,N_i,{\widehat V})~
\theta({\widehat V})
\end{equation}
with available volume 
${\widehat V}~\equiv~V~-~\sum_{i=1}^{h} v_i N_i~$.  The Laplace 
transform of Eq.~(\ref{gcparth}) gives 
\begin{eqnarray}\label{vdweosh}
p(T,\mu_1,..., \mu_h)~ & \equiv & ~
 T\lim_{V\rightarrow \infty}
\nonumber \frac{\ln{\cal Z}(T,\mu_1,...,\mu_h,V)}{V}=
p^{id}(T,\tilde{\mu}_1,...,\tilde{\mu}_h) \\[-0.1cm]
 & & \\[-0.1cm]
\nonumber &=& \sum_{i=1}^h p_i^{id}(T,\tilde{\mu}_i)~; 
\end{eqnarray}
\begin{equation}\label{muitilde}
\tilde{\mu}_i~ \equiv~ \mu_i ~ - ~v_i~ p(T,\mu_1,..., \mu_h)~,~~~
i=1,...,h~. 
\end{equation}
Particle number density for the $i$-th species can 
be calculated from Eqs.~(21,22) and is found to be 
\begin{equation}\label{partdenh}
n_i(T,\mu_1,...,\mu_h)~\equiv~
\left(\frac{\partial p}{\partial \mu_i}
\right)_{T,\mu_1,...,\mu_{i-1},\mu_{i+1},...,\mu_h}~
=~\frac{n_i^{id}(T, \tilde{\mu}_i)}
{1~+~\sum_{j=1}^h
v_j~n_j^{id}(T, \tilde{\mu}_i)}~.  
\end{equation}
The total particle number density is the sum of the partial 
values $n_i$ from $i=1$ to $i=h$.  Note that particle number 
density $n_i$ depends on all proper volume parameters 
$v_1,...,v_h$.  The partial pressures 
$p_i = p_i^{id}(T,\tilde{\mu}_i)$ are reduced to the ideal gas 
ones $p_i^{id}(T,\mu_i)$ in the limit $v_i\rightarrow 0$.  This, 
however, does not take place for $n_i$ because the particle 
number density $n_i$ with $v_i=0$ still feels the presence of 
other particles with $v_j\neq 0$ due to the suppression factor 
$[1+\sum_{j=1}^h v_j~n_i^{id}(T,{\mu}_i)]^{-1}$ in 
Eq.~(\ref{partdenh}).  

Moreover, the total entropy and energy densities 
of the VDW hadron gas are given as 
\begin{equation}\label{entrdenh}
s(T,\mu_1,...,\mu_h)~\equiv~
\left(\frac{\partial p}{\partial T}
\right)_{\mu_1,...,\mu_h}~=~\frac{\sum_{i=1}^{h}
s_i^{id}(T, \tilde{\mu}_i)}
{1~+~\sum_{j=1}^h
v_j~n_j^{id}(T, \tilde{\mu}_j)}~,
\end{equation}
\begin{equation}\label{epsdenh}
\varepsilon (T,\mu_1,...,\mu_h)~\equiv~ Ts~-~p~+~\sum_{i=1}^{h}
\mu_i n_i ~
=~\frac{\sum_{i=1}^{h}\varepsilon_i^{id}(T,
\tilde{\mu}_i)}
{1~+~\sum_{j=1}^h
v_j~n_j^{id}(T, \tilde{\mu}_j)}.
\end{equation}

\section{ Comparison with other ``Thermal Models'' }

We recapitulate in last Section a thermodynamically consistent 
formulation of the VDW repulsion in the grand canonical 
ensemble.  The problem appeared to be not trivial and many 
thermal model formulations of the VDW ``corrections'' used in 
the literature do not meet the requirement of self-consistency.  
The essential difference between the model that we use and those 
of Refs.~[4--7] is the modification of the hadron chemical 
potentials according to Eq.~(\ref{muitilde}).  The VDW 
``correction'' of the ideal gas formulae by the suppression 
factor $[1+\sum_{j=1}^h v_j~n_j^{id}(T,{\mu}_j)]^{-1}$ was 
postulated in Refs.~[4--7] for all thermodynamical functions 
including the pressure.  These ``corrected'' functions, however, 
{\em do not} satisfy fundamental thermodynamical relations.  
Besides, the consistent formulation allows the dependence on the 
excluded volumes $v_{i}$'s for particle number ratios, while in 
the formulation of Refs.~[4--7], particle ratios are always the 
same as in the ideal gas for any choices of $v_i$ values because 
the suppression factor in the denominator is identical for each 
$i$-th particle and therefore cancel each other out.  

In Ref.~\cite{braun95b}, the VDW ``correction'' 
was postulated in the form of 
\begin{equation}\label{braun}
p_{i}(T,\mu_i)~=~\frac{p^{id}_i(T,\mu_i)}
 {1+\sum_{j=1}^h
v_j~n_j^{id}(T,\mu_j)}.  
\end{equation}
The total pressure $p$ is just the sum of above partial pressures 
over all particle species $i$.  As we have shown in the previous 
Section, the VDW formulation leads to the transcendental 
equations (\ref{vdweosh},\ref{muitilde}) for the pressure 
function.  To compare it with ansatz of Eq.~(\ref{braun}), let us 
make again the Boltzmann approximation in Eq.~(\ref{vdweosh}).  
Using Eqs.~(5) and (\ref{idpress}) one finds from 
Eq.~(\ref{vdweosh}) that 
\begin{equation}\label{boltzh}
p(T,\mu_1,...,\mu_h)~\cong ~\sum_{i=1}^{h}~\exp\left(~-~
\frac{v_i~p}{T}\right)~
p^{id}_i(T,\mu_i)
\end{equation}
Assuming $v_i~p/T \ll 1$ so that 
$\exp\left(- v_i~p/T\right)~\approx (1+v_i~p/T)^{-1}$, we use 
then the expansion 
$$p(T,\mu_1,...,\mu_h)~\cong~p^{id}(T,\mu_1,...,\mu_h)~+~O(v_i)$$ 
and the relation 
\begin{equation}\label{pidnid}
p^{id}(T,\mu_1,...,\mu_h)~=~\sum_{i=1}^{h}p_i^{id}(T,\mu_i)~\cong~
 T~\sum_{i=1}^{h}n_i^{id}(T,\mu_i)
\end{equation}
(the second equality in Eq.~(\ref{pidnid}) follows because of the 
Boltzmann approximation) to find finally 
\begin{equation}\label{vzerolim}
p_i(T,\mu_i)~\approx~
\frac{p^{id}_i(T,\mu_i)}{1~+~v_i~\sum_{j=1}^h n^{id}_j(T,\mu_j)}. 
\end{equation}
Eq.~(\ref{vzerolim}) is still rather different from the 
prescription of Eq.~(\ref{braun}) used in Ref.~\cite{braun95b}.  
In Eq.~(\ref{vzerolim}) $p_i = p_i^{id}$ when $v_i = 0$.  This 
is generally true as can be clearly seen in the last equality 
of Eq.~(\ref{vdweosh}).  However, this does not take place in 
Eq.~(\ref{braun}) if some other $v_j \neq 0$.  We conclude 
therefore that the ``excluded volume correction'' in 
Eq.~(\ref{braun}) \cite{braun95b} has nothing to do with VDW 
excluded volume procedure even in the limit when all $v_i$'s 
are small, i.e., in the first order expansion over $v_i$.

\section{ Particle Number Ratios at AGS and SPS Energies }

As mentioned before, particle chemical potentials $\mu_i$ 
regulate the values of conserved charges.  For simplicity we 
neglect the effects of non-zero electrical chemical potential 
which were considered in Ref.~\cite{goryang91}.  Electrical 
chemical potential is responsible, for example, for 
$\pi^+\pi^-$-asymmetry when colliding ions are heavy and 
therefore have isotopic asymmetry, i.e., their number of 
neutrons is larger than number of protons.  These interesting 
effects are, however, not large numerically at AGS and 
especially at SPS energies. 

The chemical potential of the $i$-th particle can be written as 
\begin{equation}\label{mu}
\mu_i ~=~ b_i\,\mu_b~+~s_i\,\mu_s~, 
\end{equation}
in terms of baryonic chemical potential $\mu_b$ and strange 
chemical potential $\mu_s$, where $b_i$ and $s_i$ are the 
corresponding baryonic number and strangeness of the $i$-th 
particle.  The hadronic gas state is defined by two independent 
thermodynamical parameters, $T$ and $\mu_b$.  The strange 
chemical potential $\mu_s(T,\mu_{b})$ is determined from the 
requirement of zero strangeness 
\begin{equation}\label{strange}
n_{S}(T,\mu_b,\mu_{s})~\equiv~\sum_{i=1}^{h}
s_{i}n_{i}(T,\mu_{i})~=~0~. 
\end{equation}
In addition to  $T$ and $\mu_b$, the thermodynamical 
functions also depend on the excluded volume parameters 
$v_i$ because of Eq.~(\ref{muitilde}).  

Baryon and meson resonances and their subsequent decays to 
observed hadrons are of great importance for the measured 
particle number ratios at AGS and SPS energies.  All known 
resonance states with mass up to 2 GeV are included in 
our calculations.  Preliminary experimental data for particle 
number ratios in the collisions of Au+Au at the BNL AGS 
(11 A$\cdot$GeV/c) and Pb+Pb at the CERN SPS 
(160 A$\cdot$GeV/c) are analyzed.  We use the compilation of 
the experimental data which were presented by J. Stachel at 
QM'96 (see Ref.~\cite{stachel} and references therein).  

The enhancement of strange particle production in A+A 
collisions has attracted special attention as a possible signal 
of the quark-gluon plasma formation \cite{raf82}.  However, the 
``thermal models'' of Refs.~[4--7] which, in fact, gives the same 
results as the ideal hadron gas model have been successful in 
reproducing many of the ratios with strange hadrons measured 
at the AGS and SPS without any additional parameter to control 
strangeness abundance.  Instead, they have a problem with pion 
abundances --- for the SPS data it has been observed [4--6] 
that a simple ideal gas model is unable, within a single set of 
freeze-out parameters, to reproduce simultaneously the strange 
particles yields and anti-baryon to baryon ratios together with 
ratios where pions are involved.  Specifically, experimental pion 
to nucleon ratio and ratios of pions to other hadrons are larger 
than the ideal gas predictions.  As pion multiplicity can be 
related to the entropy of the system, the ideal hadron gas model 
appears unable to account for the large entropy per baryon of 
the freeze-out system.  The deficiency of pions in the ideal 
hadron gas calculations has become the main problem in the 
theoretical ``thermal model'' in the interpretation of the SPS 
particle production data.  Several mechanisms 
\cite{cleym93b,cleym94,raf93} 
have been proposed to remedy this problem, but no satisfactory 
answer has yet been found.  

Our model procedure is as follows.  For given values of $T$ and 
$\mu_b$, we solve the system of equations, 
Eqs.~(\ref{vdweosh},\ref{muitilde},\ref{strange}), to find 
$p(T,\mu_b)$ and $\mu_s(T,\mu_b)$.  The values of $T$ and $\mu_b$ 
are then determined from the `best fit' to particle number ratios 
not involving pions.  These best fits to the Au+Au (AGS) and 
Pb+Pb (SPS) preliminary data are shown in Fig.~1.  The chemical 
freeze-out parameters are found to be $T\cong140$ MeV, 
$\mu_{b}\cong 590$ MeV for Au+Au AGS, and $T\cong 185$ MeV, 
$\mu_b \cong 270$ MeV for Pb+Pb SPS collisions.  

All particle number densities are calculated from 
Eq.~(\ref{partdenh}) for all known stable particles and 
resonances with mass up to 2.0 GeV.  The total production yield 
of hadron $i$ is then proportional to the sum of its thermal 
density and all possible resonance decay contributions to that 
hadron $i$: 
\begin{equation}\label{total}
n_{i}^{tot} ~ = ~ n_{i}~ +n_{i}^{dec} ~=~
n_{i}~+~\sum_{j\neq i} n_j~\alpha(j,i)~, 
\end{equation}
where $\alpha(j,i)$ is the probability (branching ratio) for 
resonance $j$ to strongly decay into hadron $i$.  

In our calculations we will examine the dependence of hadron 
ratios on particle volume parameters $v_{i}$'s.  Results shown 
in Fig.~1 correspond to the same proper volume parameter 
$v_i =v$ for all hadrons.  In this case particle number ratios 
are almost independent on the value of $v$ and are the same as in 
the ideal gas and as in Refs.~[4--7].  It occurs because of the 
relation 
\begin{equation}\label{niapprox}
n_{i}^{id}(T,\tilde{\mu}_{i})~\cong~
\exp \left(-~\frac{v~
p(T,\mu_b)}{T}\right)~n_i^{id}(T,\mu_i)~. 
\end{equation}
Note that the above equality becomes exact in the Boltzmann 
approximation.  In this case a common VDW `denominator', 
${1+\sum_{j=1}^h v_jn_j^{id}(T, \tilde{\mu}_j)}$, and a common 
`numerator', $\exp (-v~p/T)$, are canceled and 
Eq.~(\ref{partdenh}) leads to 
\begin{equation}\label{partveq}
\frac{n_i(T,\mu_b)}
{n_j(T,\mu_b)}~\cong~\frac
{n_i^{id}(T,\mu_i)}
{n_j^{id}(T,\mu_j)}~. 
\end{equation}
The value of the parameter $v$ is, however, still crucial for the 
absolute values of particle number densities as well as for all 
other thermodynamical functions of the hadron gas.  At the same 
fixed $T$ and $\mu_b$, all thermodynamical functions of the 
hadron gas are smaller than in the ideal hadron gas and strongly 
decrease with increasing $v$.  

We use quantum statistics in our calculations, but some of our 
qualitative arguments depend on the validity of the Boltzmann 
approximation.  We have checked for all particle number ratios 
that the Boltzmann approximation in the ideal hadron gas 
(i.e., $v_i =0$) gives an accurate estimate to the corresponding 
quantum statistics values: $\sim$ 1--3 $\%$ for both AGS and SPS 
chemical freeze-out parameters.  In our consideration with 
$v_i >0$, each chemical potential is shifted by $-v_i p$ and the 
Boltzmann approximation always becomes even much better.  

In the case when not all of the $v_i$'s are equal, the hadron 
volume parameters $v_{i}$ do influence  the particle number 
ratios through the modification of Eq.~(\ref{muitilde}) in the 
particle chemical potentials, required by thermodynamical 
self-consistency.  The effect is quite evident that hadrons which 
take up less space (i.e., smaller values of $v_{i}$), and hence 
smaller influence on the excluded volume, have the advantage.  
The particle number ratios of those small hadrons to larger ones 
increase in comparison with the ideal gas results.  

To solve the problem with pion multiplicities in the VDW model we 
now introduce different hard-core radiuses: $r_{\pi}$ for 
pions and $r$ for all other hadrons ($r>r_{\pi}$).  Such a 
possibility with $r_{\pi}=0$ was considered in Ref.~\cite{rit97}.  
We remind that the excluded volume parameters are 
$v_i = 4\cdot {4\pi\over 3}~ r_i^3$.  Using Eqs.~(\ref{total}) 
and (\ref{partveq}), we then find 
\begin{eqnarray}\label{pionpat}
\nonumber
\frac{n_{\pi}^{tot}}{n_i^{tot}}&\cong &\frac
{\exp(-v_{\pi}p/T)
n_{\pi}^{id}(T,\mu_{\pi}=0)~+~\sum_{j\neq\pi}
\exp(-vp/T)
n_j^{id}(T,\mu_j)\alpha(j,\pi)}
{\exp(-vp/T)
n_{i}^{id}(T,\mu_{i})~+~\sum_{j\neq i}
\exp(-vp/T)
n_j^{id}(T,\mu_j)\alpha(j,i)} \\[-0.0cm]
& & \\[-0.2cm]
\nonumber
&\equiv &\frac
{\exp(\mu^{*}_{\pi}/T)~
n_{\pi}^{id}(T,\mu_{\pi}=0)~+~\sum_{j\neq \pi}
n_j^{id}(T,\mu_j)\alpha(j,\pi)}
{n_{i}^{id}(T,\mu_{i})~+~\sum_{j\neq i}
n_j^{id}(T,\mu_j)\alpha(j,\pi)}~,
\end{eqnarray}
where 
\begin{equation}\label{mupieff}
\mu^{*}_{\pi}~\equiv~
(v~-~v_{\pi})~p~ = \frac{16\pi}{3}~(r^3- r_{\pi}^3)~
p(T,\mu_b;~r,r_{\pi})~. 
\end{equation}
In Eq.~(\ref{mupieff}) we add $r$ and $r_{\pi}$ in the arguments 
of the pressure function to remind the pressure dependence 
on particle hard-core volumes through Eqs.~(\ref{vdweosh}) and 
(\ref{muitilde}).  Eq.~(\ref{pionpat}) shows that the ratio of 
pions to any hadron $i$ is changed in comparison to the ideal 
gas calculations.  From the second equality of 
Eq.~(\ref{pionpat}) the increase of the thermal pion density 
appears to be due to an effective pion chemical potential 
$\mu_{\pi}^{*}$.  In the Boltzmann approximation, it leads just 
to the additional factor of $\exp(\mu^{*}_{\pi}/T)$ in the pion 
number density of the ideal gas.  Note that all chemical 
potentials $\mu_i$ are transformed to 
$\tilde{\mu}_i = \mu_i -v_i p$ and therefore 
$\tilde{\mu}_{\pi} = -v_{\pi}p$ becomes negative because 
$\mu_{\pi}=0$.  If $v_{\pi}$ is smaller than $v_i= v$, in the 
ratios $n_{\pi}/n_{i}$, $\mu_{\pi}^*$ looks like a positive 
pion chemical potential in the ideal gas formalism.  From this 
explanation of the origin of $\mu_{\pi}^{*}$, it is clear that 
there is no restriction on its possible values by the pion mass, 
in contrast to the ideal Bose gas where always $\mu \leq m$.  

In Figs.~2 and 3 different pion to hadron ratios are shown for 
Au+Au AGS and Pb+Pb SPS collisions, respectively.  Preliminary 
experimental values are designated by dotted lines, while our 
model results are represented by solid curves, as functions of 
$\mu_{\pi}^{*}$.  The value $\mu_{\pi}^{*}=0$ corresponds to the 
ideal gas results.  We have already demonstrated that particle 
number ratios for $v_i=0$ (the ideal hadron gas) remain the same 
as those for $v_i=v=$ constant.  For all ratios in Figs.~2 and 3 
we find that the experimental values systematically exceed the 
ideal gas results ($\mu_{\pi}^*=0$).  To fit data, one needs 
$\mu_{\pi}^{*} >0$ and from Eq.~(\ref{mupieff}) it means 
$r>r_{\pi}$.  Our $T$ and $\mu_b$ values are already fixed both 
for AGS and SPS from ratios given in Fig.~1.  Assuming $r_{\pi}$ 
different from $r_i=r$ for other hadrons we obtain no changes in 
the VDW model values for the ratios shown in Fig.~1.  As pions 
have no influence on those ratios, they remain the same as the 
ideal hadron gas results.  

At fixed $T$ and $\mu_b$ the value of $\mu_{\pi}^{*}$ 
is a complicated function of $r_{\pi}$ and $r$.  The ratios in 
Figs.~2 and 3 feel not specific values of $(r_{\pi},r)$ but 
$\mu_{\pi}^{*}$ value.  From the preliminary experimental data 
we find $\mu_{\pi}^{*}\cong 100$ MeV for Au+Au AGS collisions 
and $\mu_{\pi}^{*}\cong 180$ MeV for Pb+Pb SPS ones.  We stress 
that the problem with pion deficiency observed in Refs.~[5,6] 
for SPS energies looks similar to what we find in {\it both} AGS 
and SPS data analyzed above.  This deficiency of pions for the 
preliminary data in Pb+Pb SPS collisions is not so drastic as 
in S+Pb SPS data, where the pion to nucleon ratio is close to 
8 \cite{pionnucl} and is approximately 2 times larger than the 
ideal gas result.  To have an agreement with data in Figs.~2 and 
3 we need only about 30$\%$ larger pion to other hadrons ratios 
than those in the ideal hadron gas.  With $r_{\pi}<r$ (and 
therefore $\mu_{\pi}^{*}>0$) the thermal pion number density 
would increase by a factor of $\exp(\mu_{\pi}^{*}/T)$.  It is 
about 2 for Au+Au AGS and about 2.6 for Pb+Pb SPS collisions to 
have an agreement with data.  The effect for the pion thermal 
density is therefore quite strong.  However, it does not 
strongly alter the pion to hadron ratios, Eq.~(\ref{pionpat}) 
because at both AGS and SPS energies the pion production is 
essentially dominated by the resonance decay contributions.  
Let us also remind the possibility of the chemical 
non-equilibrium effects for pions discussed in 
Ref.~\cite{mupi}.  It would lead to the chemical potential 
$\mu_{\pi}>0$ with its values being always smaller than the pion 
mass $m_{\pi}$.  For $r_{\pi}<r$, we have obtained in the VDW 
model the effective pion chemical potential $\mu_{\pi}^{*}$ 
whose value is not restricted by the pion mass, and no chemical 
non-equilibrium effects are required.

\section{ Particle Hard-Core Radii }

Our fits to the preliminary data of Au+Au (AGS) and Pb+Pb (SPS) 
on the particle number ratios with VDW hadron gas model are 
shown in Figs.~1--3.  From fitting the data we have found the 
following model parameters: 
\begin{equation}\label{AGS}
\mbox{AGS}:~~~T~\cong~140~\mbox{MeV}~,~~\mu_b~\cong~590~
\mbox{MeV}~,~~\mu^{*}_{\pi}~\cong ~100~\mbox{MeV}~, 
\end{equation}
\begin{equation}\label{SPS}
\mbox{SPS}:~~~T~\cong~185~\mbox{MeV}~,~~\mu_b~\cong~270~
\mbox{MeV}~,~~\mu^{*}_{\pi}~\cong ~180~\mbox{MeV}~. 
\end{equation}
The set of all possible values of $r_{\pi}$ and $r$ which give 
the same value of $\mu_{\pi}^{*}$ in Eq.~(\ref{mupieff}) for the 
AGS or SPS defines a curve in $(r_{\pi},r)$-plane.  The two 
curves corresponding to the AGS and SPS respectively are shown 
in Fig.~4, together with their intersection point 
($r_{\pi}\cong 0.62$ fm, $r \cong 0.8$ fm).  This intersection 
point is a solution for $r_{\pi}$ and $r$ in the VDW hadron gas 
model to fit simultaneously the AGS and SPS data for all 
particle number ratios.  

In Tables I and II, we show the values of the total meson number 
density $n_m$, baryon number density $n_b$, total particle number 
density $n_{tot}$ (antibaryons included), total energy density 
$\varepsilon$ and total pion number density $n_{\pi}^{tot}$, 
which includes both thermal pions and contributions from 
resonance decays, at chemical freezeout.  The ideal hadron gas 
($r_{\pi}=r=0$) results are given in the first row, and those 
from our calculations with different $(r_{\pi},r)$ values along 
the curves in Fig.~4 are shown in the remaining 4 rows.  We 
emphasize that the last rows in the Tables give our results for 
the intersection point ($r_{\pi}\cong 0.62$ fm, $r\cong 0.8$ fm), 
and that these values are our predictions within the VDW hadron 
gas model for the chemical freezeout at Au+Au AGS and Pb+Pb 
SPS collisions, respectively.  We believe that these $r_{\pi},r$ 
values, along with the corresponding densities, give a reasonable 
physical solution for the chemical freeze-out state in A+A 
collisions considered.

\section{ Summary }

A self-consistent hadron gas model with the VDW excluded volume 
is considered and critically compared with other 
``thermal models'' used in the literature.  This approach is 
then adopted to analyze the preliminary data of Au+Au (AGS) and 
Pb+Pb (SPS).  Within the VDW model, the obtained values of 
particle number and energy densities, and other thermodynamical 
functions at chemical freezeout are very different from those 
obtained in the ideal hadron gas, as seen in the first and last 
rows in Tables I and II.  Because of their strong effects on 
hadron thermodynamical functions, the VDW gas formulation should 
be properly treated.  

The preliminary data of Au+Au (AGS) and Pb+Pb (SPS) for the 
particle number ratios can be fitted in the VDW hadron gas 
as shown in Figs.~1--3.  The model parameters, given in 
Eqs.~(\ref{AGS},\ref{SPS}), lead to the enhancement of 
pions in pion to hadron ratios as compared to the ideal hadron 
gas model predictions.  This enhancement in pions is regulated 
by $\mu_{\pi}^*$ ($>0$) and is explained in the VDW model by 
a smaller pion ``hard-core radius'' than those of all other 
hadrons.  

The obtained parameters for Au+Au (AGS) and Pb+Pb (SPS) define 
two curves in the $(r_{\pi},r)$-plane, respectively.  These two 
curves shown in Fig.~4 intersect at the point 
($r_{\pi}\cong 0.62$ fm, $r \cong 0.8$ fm).  It is the solution 
in the VDW hadron gas model to fit simultaneously the AGS and 
SPS data for all particle number ratios.  The absolute values of 
the particle number and energy densities in the VDW hadron gas 
for this solution are listed in the last rows of Tables I and 
II.  We stress that these values are much smaller than those in 
the ideal hadron gas at the same $T$ and $\mu_b$ shown in the 
first rows of Tables I and II.  There is an experimental 
estimate for the freeze-out pion number density in Si+Pb central 
collisions at the AGS energy: 
$n_{\pi}^{exp}\cong 0.063$ fm$^{-3}$ \cite{bar}.  Our result 
$n_{\pi}^{tot}\cong 0.054$ fm$^{-3}$ for Au+Au AGS collisions 
shown in the last row of the Table I is quite close to this 
experimental estimate.  

The next step is naturally to fit hadron momentum spectra.  It 
requires the inclusion of the longitudinal and transverse 
collective flow effects.  The temperature $T\cong 185$ MeV looks 
a little too high for use in the fitting of the transverse 
momentum spectra in Pb+Pb (SPS) collisions.  For example, the 
pion inverse slope parameter is near $190$ MeV \cite{jone} and 
the freeze-out temperature $T\cong 185$ MeV seems to leave almost 
no room for the transverse collective motion effects.  However, 
two facts should be taken into account.  First, the temperature 
determined from particle number ratios is for the chemical 
freezeout which could be higher than the thermal freeze-out 
temperature used in particle spectra calculations.  Second, we 
remind again that there are large resonance decay contributions 
to pion production.  Even after the enhancement of thermal pions 
with $\mu_{\pi}^*\cong 180$ MeV, resonance decays contribute more 
than 60$\%$ to the final pions.  These resonance decays are known 
to lead to a lower pion `effective temperature' (the inverse 
slope parameter) at the transverse pion mass less than 1 GeV.

\acknowledgements

This work is supported in part by the National Science 
Council of Taiwan under grant Nos.\ NSC 86-2112-M-001-010, 
86-2112-M-002-016.  MIG gratefully acknowledges the financial 
support by NSC, and the kind hospitality of the Physics 
Department, National Taiwan University. He is also thankful 
to St.~Mr\'owczy\'nski for reading the manuscript and useful 
comments.

\begin{table}
\caption{ Meson number density $n_m$, baryon number density 
$n_b$, total particle number density $n_{tot}$, energy density 
$\varepsilon$ and total pion number density $n_{\pi}^{tot}$, 
which includes thermal pions and contributions from decays, at 
the freezeout for AGS Au+Au collision at 11 A$\cdot$GeV/c.  }
\begin{tabular}{cccccc}
($r_{\pi}$, $r$) [fm] & $n_m$ [fm$^{-3}$] & $n_b$ [fm$^{-3}$] & 
$n_{tot}$ [fm$^{-3}$] & $\varepsilon$ [GeV/fm$^3$] & 
$n_{\pi}^{tot}$ [fm$^{-3}$] \\
\tableline
(0.00, 0.00) & 0.200 & 0.402 & 0.603 & 0.722 & 0.400 \\
(0.00, 0.50) & 0.097 & 0.130 & 0.227 & 0.248 & 0.161 \\
(0.20, 0.52) & 0.089 & 0.120 & 0.209 & 0.228 & 0.149 \\
(0.40, 0.61) & 0.061 & 0.084 & 0.145 & 0.159 & 0.103 \\
(0.62, 0.80) & 0.032 & 0.043 & 0.075 & 0.083 & 0.054 
\end{tabular}
\end{table}


\begin{table}
\caption{ Meson number density $n_m$, baryon number density 
$n_b$, total particle number density $n_{tot}$, energy density 
$\varepsilon$ and total pion number density $n_{\pi}^{tot}$, 
which includes thermal pions and contributions from decays, at 
the freezeout for SPS Pb+Pb collision at 160 A$\cdot$GeV/c.  }
\begin{tabular}{cccccc}
($r_{\pi}$, $r$) [fm] & $n_m$ [fm$^{-3}$] & $n_b$ [fm$^{-3}$] & 
$n_{tot}$ [fm$^{-3}$] & $\varepsilon$ [GeV/fm$^3$] & 
$n_{\pi}^{tot}$ [fm$^{-3}$] \\
\tableline
(0.00, 0.00) & 0.771 & 0.404 & 1.251 & 1.585 & 1.213 \\
(0.00, 0.46) & 0.278 & 0.095 & 0.391 & 0.429 & 0.382 \\
(0.20, 0.48) & 0.244 & 0.084 & 0.344 & 0.378 & 0.336 \\
(0.40, 0.59) & 0.139 & 0.049 & 0.197 & 0.219 & 0.193 \\
(0.62, 0.80) & 0.064 & 0.022 & 0.090 & 0.100 & 0.088 
\end{tabular}
\end{table}

\figure{ Fig.~1: 
Points are the preliminary experimental data for the particle 
number ratios (see Ref.~\cite{stachel} and references therein) 
for Au+Au AGS and Pb+Pb SPS collisions (in the lower and upper 
part of the figure respectively).  The short horizontal lines 
are the model fit with $T\cong 140$ MeV, $\mu_b \cong 590$ MeV 
(AGS), and $T\cong 185$ MeV, $\mu_b \cong 270$ MeV (SPS).  }

\figure{ Fig.~2: 
Pion to hadron ratios for Au+Au AGS collisions.  The 
experimental data (see Ref.~\cite{stachel} and references 
therein) are shown by the dotted horizontal lines.  The solid 
lines are the VDW hadron gas model results as the functions of 
$\mu^{*}_{\pi}$ (see the text for details).  An agreement with 
data corresponds to $\mu^{*}_{\pi}\cong 100$ MeV.  }

\figure{ Fig.~3: 
Pion to hadron ratios for Pb+Pb SPS collisions.  The 
experimental data (see Ref.~\cite{stachel} and references 
therein) are shown by the dotted horizontal lines.  The solid 
lines are the VDW hadron gas model results as the functions of 
$\mu^{*}_{\pi}$ (see the text for details).  An agreement with 
data corresponds to $\mu^{*}_{\pi}\cong 180$ MeV.  }

\figure{ Fig.~4: 
The solutions of the equations $\mu_{\pi}^* =$ constant 
for AGS parameters, Eq.~(\ref{AGS}), (the dashed curve) 
and SPS ones, Eq.~(\ref{SPS}), (the dotted curve) shown in 
$(r_{\pi},r)$-plane.  The intersection point is approximately at 
($r_{\pi}\cong 0.62$ fm, $r\cong 0.8$ fm).  }


\begin{references}

\bibitem{qm96} Proceedings of Quark Matter 96 in Nucl.\ Phys.\ 
{\bf A610} (1996).  

\bibitem{gor91} D. H. Rischke, M. I. Gorenstein, H. St\"{o}cker 
and W. Greiner, Z. Phys.\ C {\bf 51}, 485 (1991).  

\bibitem{gor93} J. Cleymans, M. I. Gorenstein, J. Stalnacke and 
E. Suhonen, Phys.\ Scripta {\bf 48}, 277 (1993).  

\bibitem{cleym93a} J. Cleymans and H. Satz, Z. Phys.\ C {\bf 57}, 
135 (1993).  

\bibitem{cleym93b} J. Cleymans, K. Redlich, H. Satz and E. 
Suhonen, Z. Phys.\ C {\bf 58}, (1993) 347.  

\bibitem{cleym94} J. Cleymans, K. Redlich, H. Satz and E. 
Suhonen, Nucl.\ Phys.\ {\bf A566}, 391c (1994).  

\bibitem{barun95a} P. Braun-Munziger, J. Stachel, J. P. Wessels 
and N. Xu, Phys.\ Lett.\ B {\bf 344}, 43 (1995).  

\bibitem{braun95b} P. Braun-Munziger, J. Stachel, J. P. Wessels 
and N. Xu, Phys.\ Lett.\ B {\bf 365}, 1 (1995).  

\bibitem{lan75} L. D. Landau and E. M. Lifshitz, 
{\it Statistical Physics}, (Pergamon, 1975).  

\bibitem{goryang91} M. I. Gorenstein and S. N. Yang, 
Phys.\ Rev.\ C {\bf 44}, 2875 (1991); M. I. Gorenstein, 
H. G. Miller, R. M. Quick and S. N. Yang, Phys.\ Rev.\ C 
{\bf 50}, 2232 (1994);  M. I. Gorenstein and H. G. Miller, 
Phys.\ Rev.\ C {\bf 55}, 2002 (1997).  

\bibitem{stachel} J. Stachel, Nucl.\ Phys.\ {\bf A610}, 509c 
(1996).  

\bibitem{raf82} J. Rafelski, Phys.\ Rep.\ {\bf 88}, 331 (1982); 
P. Koch, B. M\"uller and J. Rafelski, Phys.\ Rep.\ {\bf 142}, 167 
(1986).  

\bibitem{raf93} J. Letessier, A. Tounsi, U. Heinz, J. Solfrank 
and J. Rafelski, Phys.\ Rev.\ Lett.\ {\bf 70}, 3530 (1993); 
Phys.\ Rev.\ D {\bf 51}, 3408 (1995).  

\bibitem{rit97} R. A. Ritchie, M. I. Gorenstein and 
H. G. Miller, Z. Phys.\ C (1997) (in print).  

\bibitem{pionnucl} A. Iyono et al., EMU05 Coll., Nucl.\ Phys.\ 
{\bf A544}, (1992) 455c; R. Holy\'nski, Nucl.\ Phys.\ {\bf A556}, 
191c (1994).  

\bibitem{mupi} M. Kataja and P. V. Ruuskanen, Phys.\ Lett.\ B 
{\bf 243}, 181 (1990); P. Gerber, H. Leutwyler and J. L. Goity, 
Phys.\ Lett.\ B {\bf 246}, 513 (1990).  

\bibitem{bar} J. Barrette at al., E814 Coll., Phys. Lett.\ B 
{\bf 333}, 33 (1994).  

\bibitem{jone} P. G. Jones and the NA49 Collaboration, 
Nucl.\ Phys.\ {\bf A610}, 188c (1996).  

\end{references}
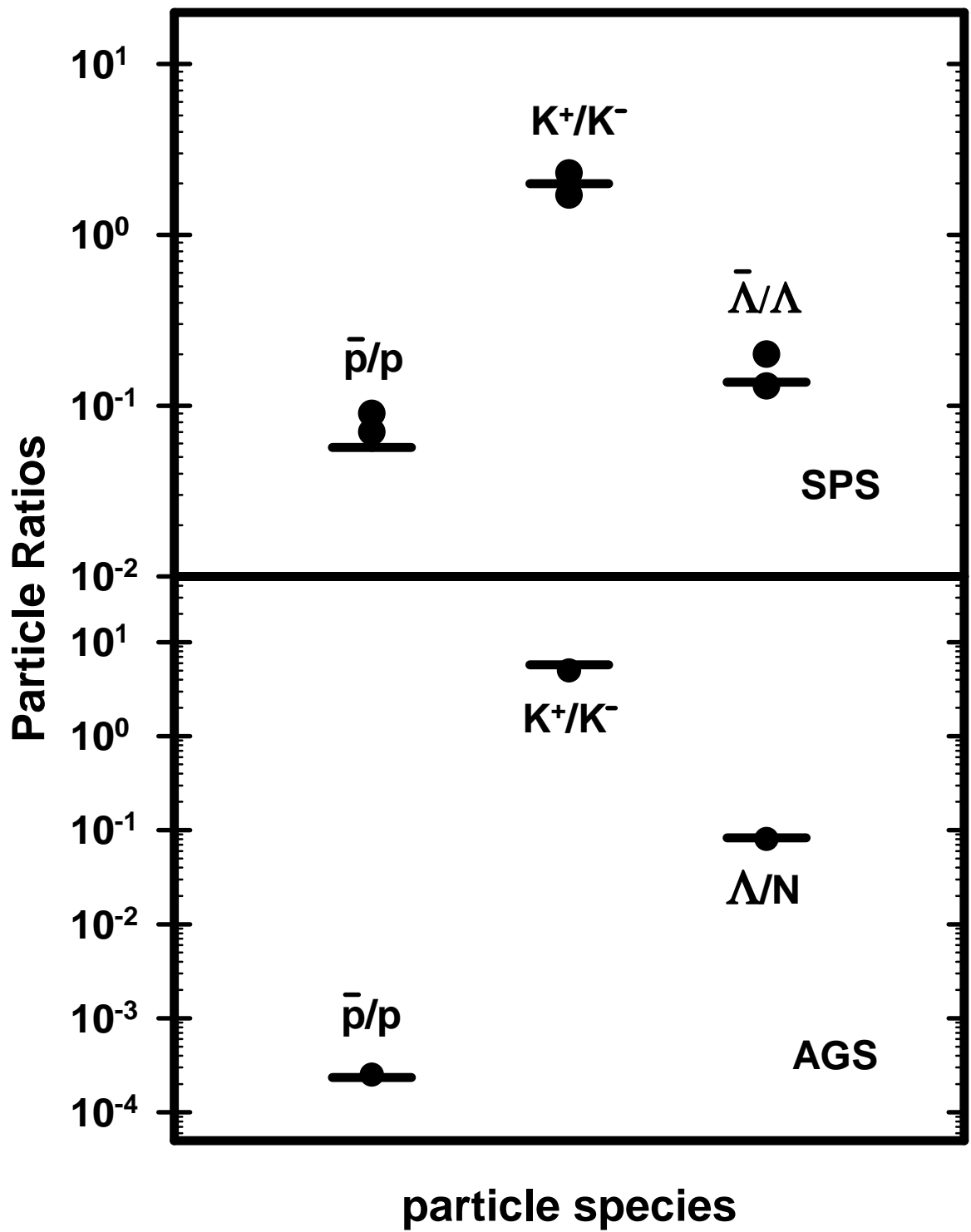
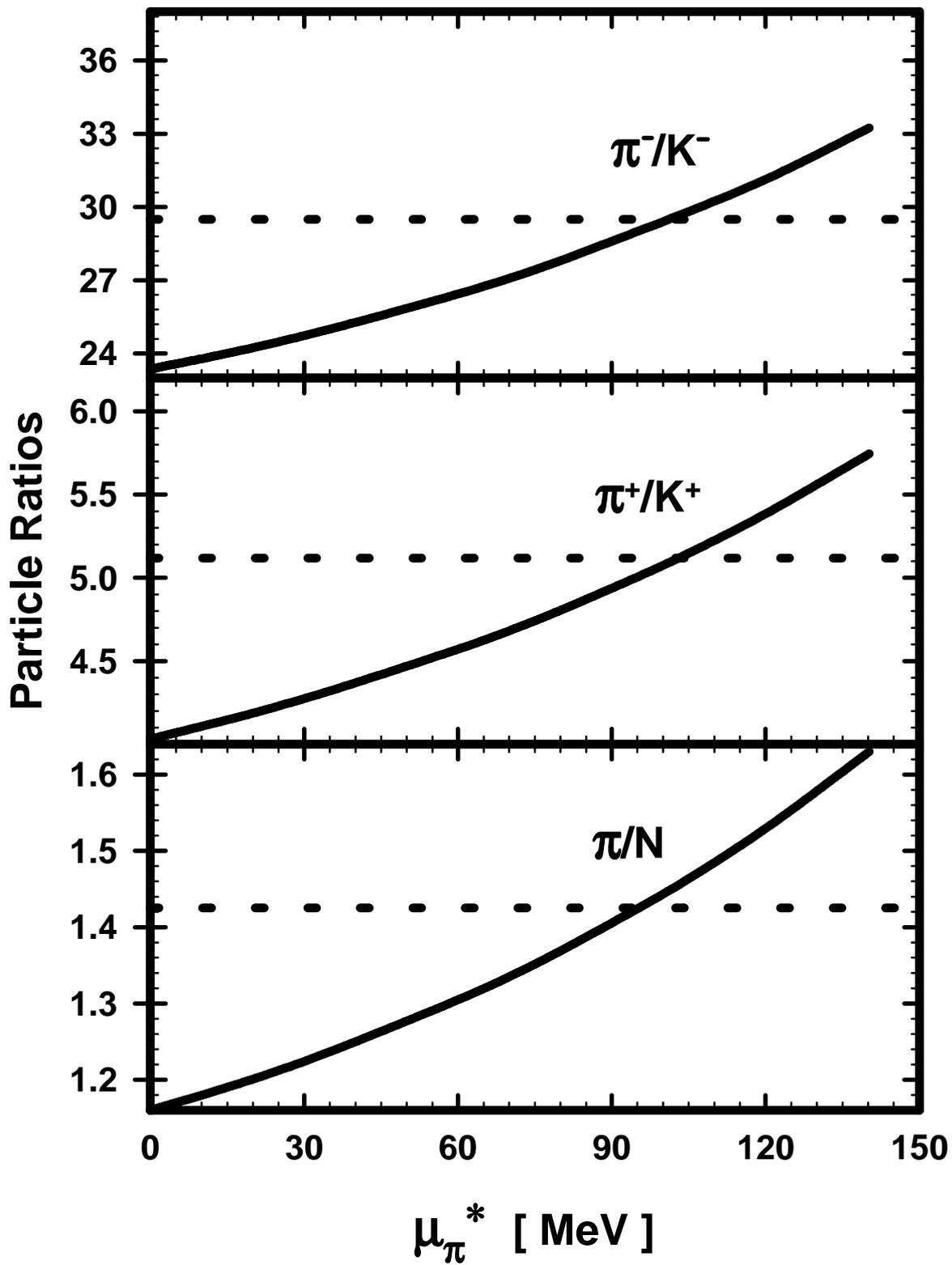
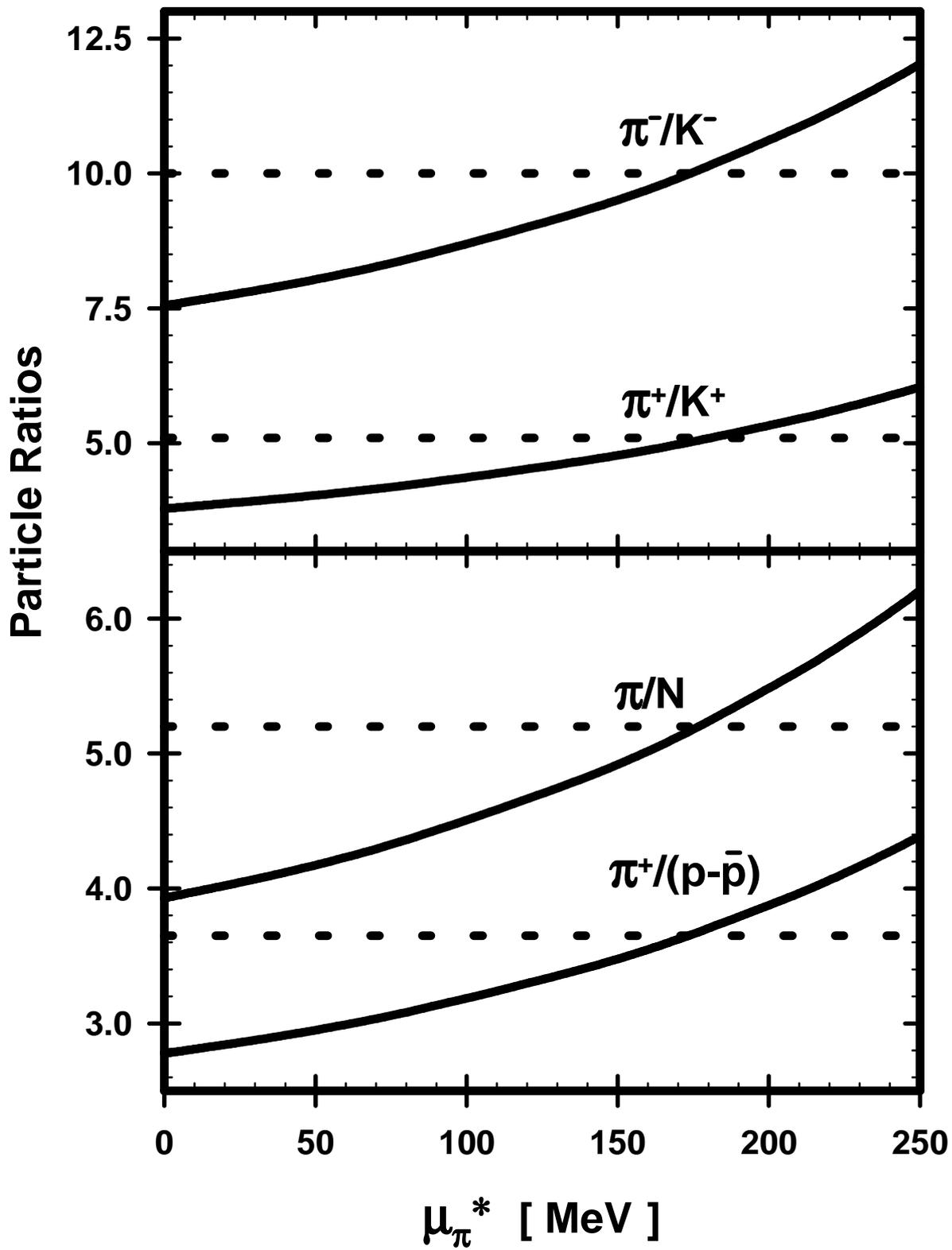
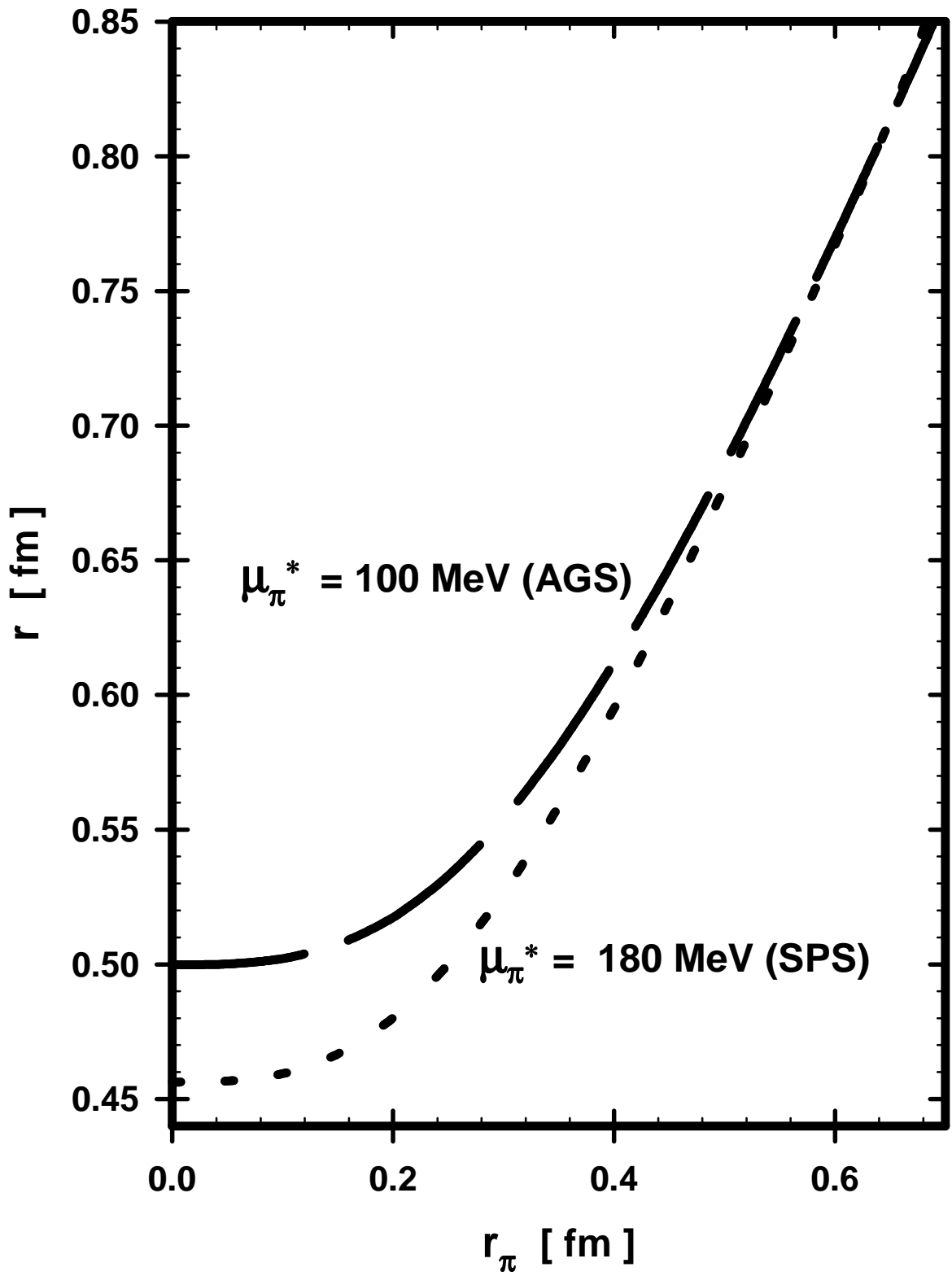
\end{document}